\documentclass[11pt,a4paper]{article}

\usepackage{graphicx}
\usepackage{physics}
\usepackage{subfigure}
\usepackage{amssymb}
\usepackage[affil-it]{authblk}
\usepackage{blindtext}

\begin{document}

\title{Vacuum-based quantum random number generator using multi-mode coherent states}



\author[1,2,*]{Samsonov E.O.}
\author[1]{Pervushin B.E.}
\author[1,2]{Ivanova A.E.}
\author[1]{Santev A.A.}
\author[1,2]{Egorov V.I.}
\author[1,2]{Kynev S.M.}
\author[3]{Gleim A.V.}

\affil[1]{ITMO University, Kronverkskiy, 49, St.Petersburg, 197101, Russia}
\affil[2]{Quanttelecom LLC., 6 Line, 59, St.Petersburg, 199178, Russia}
\affil[3]{JSCo Russian Railways, Department of Quantum Communications, Novaya Basmannaya, 2, Moscow, 107174, Russia}
\affil[*]{Corresponding author: Samsonov E.O., eosamsonov@itmo.ru}


\date{}
\maketitle
\begin{abstract}
We present an optical quantum random number generator based on vacuum fluctuation measurements that uses multi-mode coherent states generated by electro-optical phase modulation of an intense optical carrier. In this approach the weak coherent multi-mode state (or a vacuum state) interferes with the carrier, which acts as a local oscillator, on each side mode independently. The proposed setup can effectively compensate for deviations between the two arms of a balanced detector by controlling the modulation index of the  electro-optical phase modulator. We perform a proof-of-principle experiment and demonstrate random number generation with a possibility of real-time randomness extraction at the rate of 400 Mbit/s.
\end{abstract}

\section{Introduction}
Random number generation is highly relevant for many branches of science and technology \cite{Ferrenberg,Ulam1949,gennaro2006,Tittel2002,brunner}. Common solutions are divided into two groups: pseudo \cite{nisan1994} and physical \cite{johnston2018random} random number generators (RNG). The pseudo RNGs rely on deterministic processes, and their outcome may be predictable under certain conditions. Physical RNGs that utilize the measurement outcomes in complex classical systems are also predictable, in a sense, due to the intrinsic deterministic nature of any non-quantum natural system \cite{haw2015}. These types of generators can satisfy the needs of most applications, however, they are unsuitable for fields that require "fundamentally true" randomness, such as classical and quantum cryptography. It should also  be noted that in practice certified non-quantum physical RNGs are relatively slow for modern applications. For instance, "true" RNGs with rates from hundreds of megahertz to gigahertz are required for running contemporary quantum cryptography systems, especially taking into account the need to accumulate large bit statistics to eliminate the finite size effects in the quantum keys \cite{Renner2008,kozubov2019finite}.

One of the promising approaches offering high-speed true randomness relies on using entropy sources based on quantum optics effects. The non-determinism in this quantum random number generation (QRNG) is ensured by the fundamentally probabilistic nature of quantum processes. Practical QRNGs can be implemented by registering single-photons in different optical modes \cite{jennewein2000,cao2016,martin2015}, using entangled photons \cite{pironio2010, Xu2016}, phase noise of lasers \cite{Guo2010, xu2012,abellan2014}, or measurements of the quantum fluctuations of vacuum states \cite{Shi2016,Gabriel2010,Zhu2012,zhou2019,xu2019,zheng2019,avesani2018,huang2019}. 

Vacuum-based QRNGs are extensively developed thanks to the convenience of state preparation, insensitivity to detection efficiency, relativity compact optical setup and high generation rates. Parallel vacuum-based QRNG with high real-time generation rate \cite{Guo2019,haylock2019}, as well as the ones integrated onto a photonic chip \cite{Raffaelli2018} have been proposed. Vacuum-based QRNGs generally use homodyne detection schemes that measure the quadrature amplitudes of vacuum states. By definition, homodyne detection is performed using interference of a weak signal (the vacuum state in case of QRNG) with an intense signal (local oscillator) on a $50/50$ beam splitter, followed by a balanced detector. After the interference, photon number difference between the two arms of the balanced detector and, correspondingly, the difference in the number of photo-electrons is proportional to the measured amplitude of the vacuum state.

We propose a new approach to implementing a detection scheme for measuring the quantum fluctuations of vacuum states. It is based on the methodology used in subcarrier wave quantum key distribution systems (SCW QKD), where the signal photons are not emitted directly by a laser source but are generated on subcarrier frequencies, or sidebands, in course of phase modulation of an intense optical carrier wave \cite{merolla1999,gleim2016,miroshnichenko2018,kynev2017,chistiakov2019}. Earlier it was shown that in SCW QKD one can implement a coherent detection scheme, using the carrier wave as a local oscillator \cite{Samsonov2019, melnik}. The scheme is on purpose referred to as "coherent" (not "homodyne") detection, because, strictly speaking, it corresponds to a more fundamental definition. In SCW QKD scheme with coherent detection the weak coherent multimode states interfere at the phase modulator with the carrier wave on each side mode independently. Therefore, depending on the relative phase induced by the modulators in QKD sender and receiver modules, a major part of the signal will either go to the sidebands (in case of constructive interference) or remain at the carrier (in case of destructive interference) \cite{gleim2016}. Then the two output modes (the carrier and the sidebands) are detected by photodiodes, and their respective photocurrent values are subtracted \cite{Samsonov2019}. If a vacuum state is used instead of a weak coherent multimode state, the local oscillator also interferes with it, and the difference in photocurrents is proportional to the vacuum state on the sidebands. As we describe below, this approach can be used for constructing a QRNG. 

In this paper we demonstrate the possibility of implementing a vacuum-based QRNG using multi-mode coherent states and compose a mathematical model for the detection scheme. We experimentally demonstrate random number generation with real-time randomness extraction at 400 Mbit/s rate. We show that by controlling the modulation index of the phase modulator we can compensate the detection setup asymmetry. We quantify the randomness via min-entropy conditioned on classical noise using approach from \cite{haw2015, guo2018}, which allows to estimate the number of almost uniform and independent random bits that can be extracted from an untrusted source. Finally we apply DieHard \cite{marsaglia} and NIST \cite{rukhin} statistical tests to assess the randomness of our QRNG.

\section{Setup and principles of operation}

\begin{figure}[ht!]
\centering
\includegraphics[width=\linewidth]{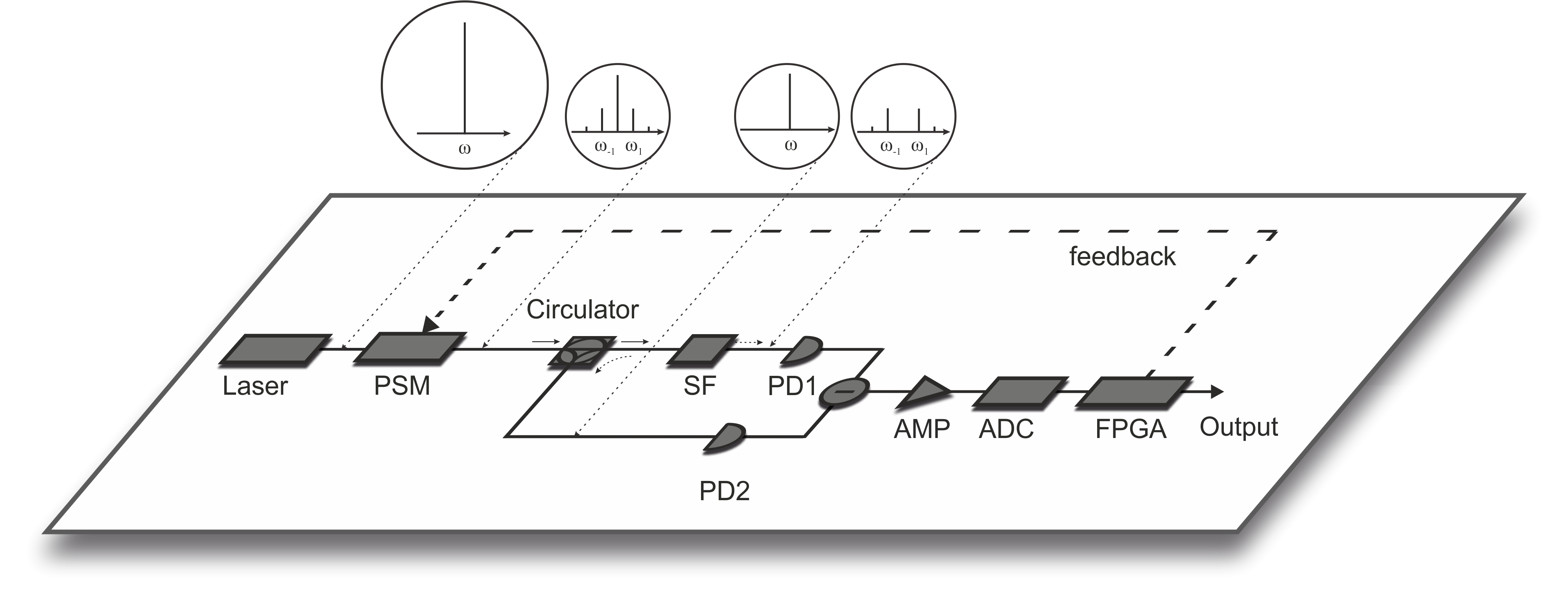}
\caption{Principal scheme of the vacuum-based random number generator using multi-mode coherent states. Diagrams in circles show the absolute values of signal amplitude spectrum taking into account only the third-order subcarriers. PSM is an electro-optical phase modulator; SF is a spectral filter that cuts off the carrier; PD is a photodiode; ADC is an analog-to-digital converter; FPGA is a field-programmable gate array.} 
\label{fig1}
\end{figure}

Figure \ref{fig1} shows the principal scheme of the proposed vacuum-based random number generator. A fiber-coupled laser emits coherent light with frequency $\omega$. The initial state can be described as $\ket{\sqrt{\mu_{0}}}_0 \otimes \ket{vac}_{SB}$, where $\ket{vac}_{SB}$ is the vacuum state on sidebands and $\ket{\sqrt{\mu_{0}}}_0$ is the carrier wave coherent state with the average number of photons $\mu_{0}$. The initial state enters a lithium niobate (LiNbO$_3$) electro-optical travelling wave phase modulator (PSM) operated by the microwave field with frequency $\Omega$. As a result of modulation, a pairs of sidebands is formed at frequencies $\omega_{k}=\omega+k\Omega$, where integer $k$ runs between the limits $-S\leq k \leq S$. The local oscillator can interfere with vacuum state on each side mode independently. Hence, we acquire a multi-mode coherent state 

\begin{equation}
|\psi\rangle = \bigotimes_{k=-S}^S|{\alpha_k}\rangle_k,
\end{equation}
with coherent amplitudes
\begin{equation}\label{alpha}
\alpha_k=\sqrt{\mu_0}d^S_{0k}(\beta)e^{-ik\theta},
\end{equation}
where $\theta$ is a constant phase and $d^S_{nk}(\beta)$ is the Wigner d-function \cite{Varshalovich1988}. Argument of the d-function $\beta$ is determined by the modulation index $m$, disregarding the modulator medium dispersion this dependence can be written as
\begin{equation}\label{beta}
\cos{({\beta})}=1-\frac{1}{2}{\left(\frac{m}{S+0.5}\right)^2}.
\end{equation}
After that a narrow spectral filter and an optical circulator separate the carrier from the sidebands, and the average number of photons arriving at the first arm of the balanced detector is

\begin{equation}
\label{eq:n1}
n_{1}\left(\beta\right)= \vartheta|a_0|^2 + \sum_{k\neq{0}}|a_k|^2 =\mu_{0} \eta_{SB} \left(1-(1-\vartheta)\left|d_{00}^{S}\left(\beta\right)\right|^{2}\right),
\end{equation}
Optical sidebands losses in the detection module can be described by a transmittance coefficient $\eta_{SB}$, and $\vartheta$ is carrier wave extinction factor (defines the carrier fraction that passes the spectral filter). The average number of photons arriving at the second arm of the detector is
\begin{equation}
\label{eq:n2}
n_{2}\left(\beta\right)=\mu_{0} \eta_C (1-\vartheta)\left|d_{00}^{S}\left(\beta\right)\right|^{2},
\end{equation}
where $\eta_{C}$ is the carrier transmittance coefficient. The detailed description of electro-optical modulation process for the quantum states can be found in \cite{miroshnichenkoj}.

Optical power in the two arms is detected on respective photodiodes, the photocurrents of which are then subtracted. The photon number in the arms can be controlled by varying the modulation index, that should be chosen so that the amplitude ratio between the carrier and all the subcarrier waves equals unity i.e $n_1-n_2 = 0$. In this case the photo-electron number difference will be proportional to the amplitude of the sidebands vacuum state. The normalized quadrature value of the signal is obtained as
\begin{equation}
    x_0 = \frac{(n_1-n_2)\cdot s}{2 \cdot\sqrt{\mu_{0}}},
\end{equation}
where $s$ is detector sensitivity. Using feedback in order to control modulation index, one can compensate the asymmetry between two arms of the generator caused by the fabrication imperfection in the practical devices. The measured values of the vacuum state are random and have Gaussian probability density distribution $p_m$ with variance $\sigma_m^2$, centered at zero 
\begin{equation}
\displaystyle
p_m(x)=\frac{1}{\sigma_m\sqrt{2\pi}}e^{-{\frac{(x-x_0)^2}{2\sigma^2_m}}}.
\end{equation}
In order to obtain random numbers we digitize the resulting distribution using an ADC and then apply a randomness extractor.

\section{Experimental results}
\begin{figure}[ht!]  
\vspace{0pt} \centering 
\subfigure[]{
\includegraphics[width=0.38\linewidth]{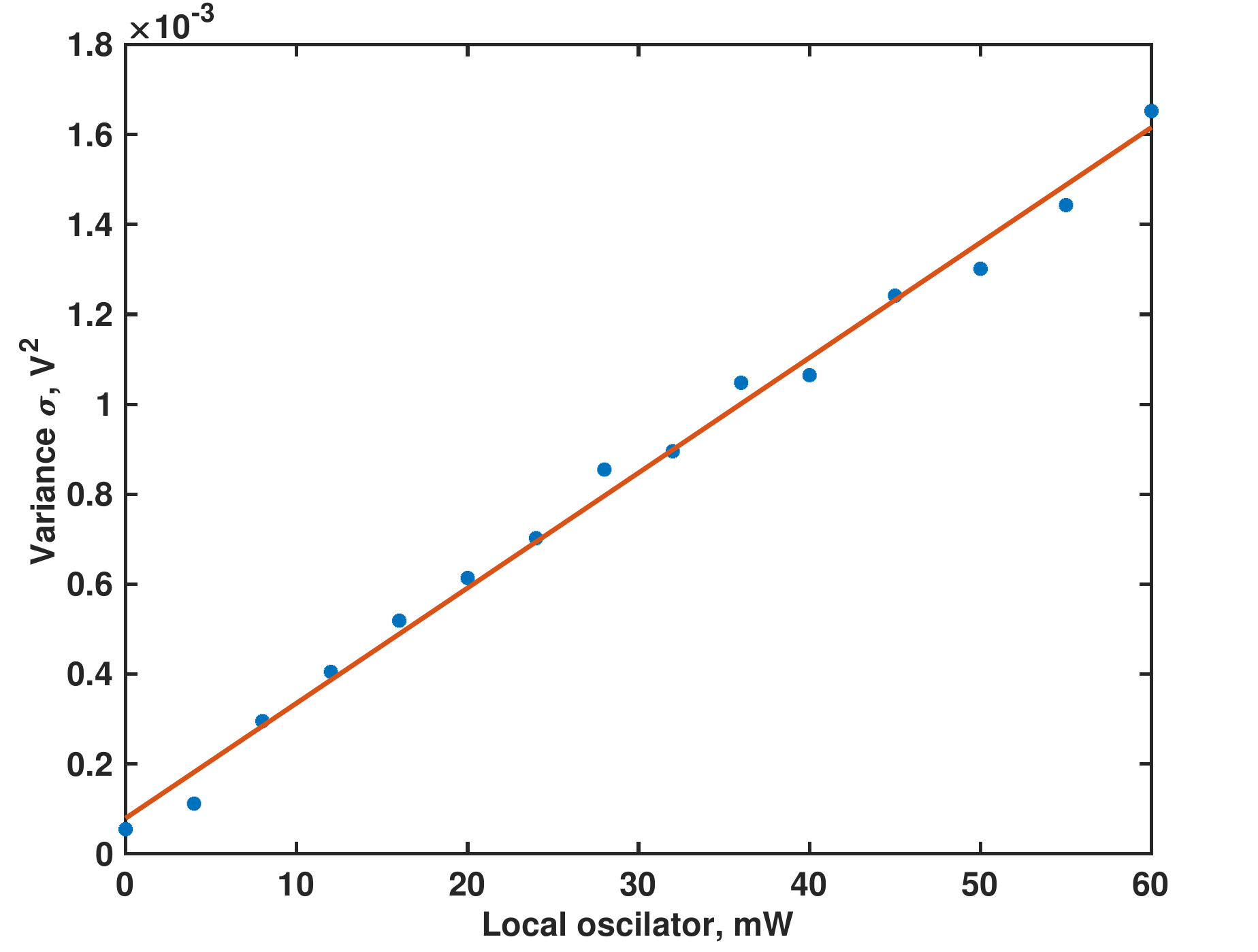}
\label{Fig:VarLO} }  
\hspace{4ex}
\subfigure[]{
\includegraphics[width=0.4\linewidth]{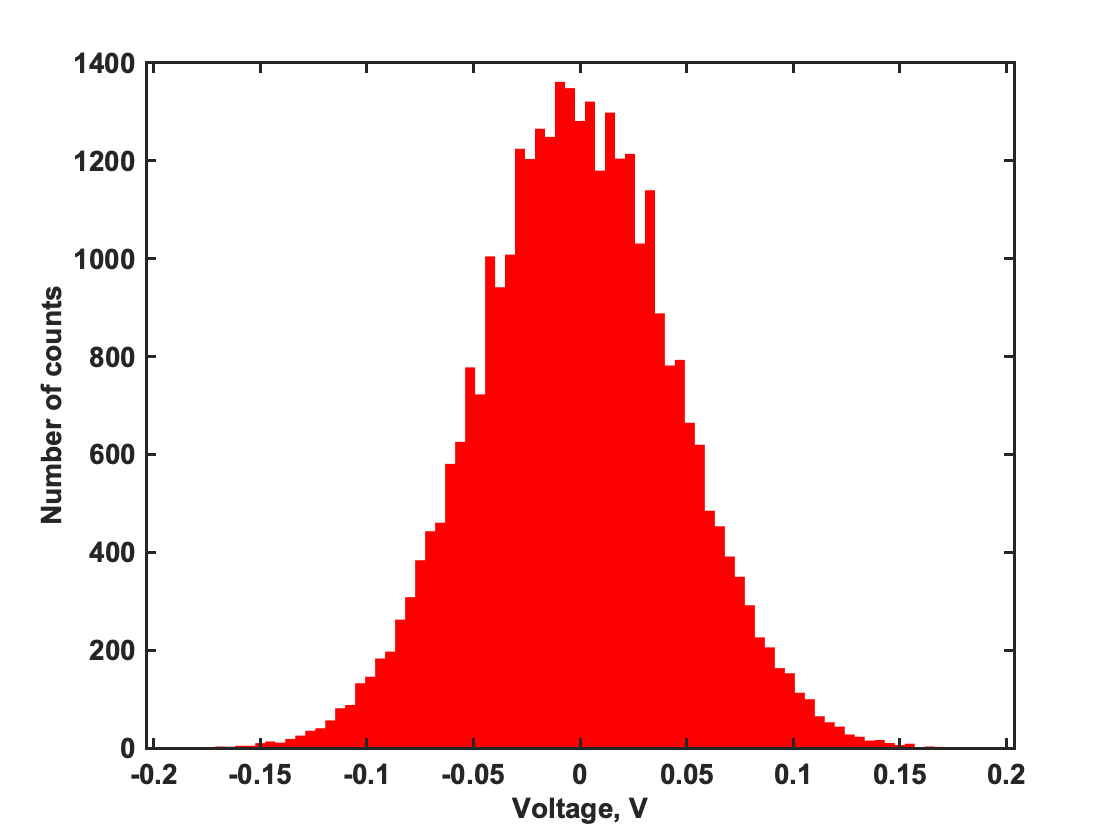} \label{Fig:Noise} }
\caption{\subref{Fig:VarLO} Voltage variance, $\sigma_{m}^2$ as a function of LO power. \subref{Fig:Noise} Experimentally obtained measurement probability distribution (based on $10^3$ measurement outcomes), LO power 40 mW.}
\end{figure}

To perform a proof-of-principle experiment we implement the setup shown in figure \ref{fig1}. A 1550 nm 40 mW fiber-coupled laser serves as the local oscillator. It is directed into the LiNbO$_3$ PSM with 3 dB loss to perform signal modulation. The system clock frequency is set by a voltage controlled oscillator (VCO). The VCO signal is send to an external phase-locked loop device, where its frequency is multiplied to generate an output electrical signal with frequency $\Omega$ = 4.2 GHz. The clock signal from the VCO also controls operation of a field-programmable gated-array (FPGA) logic module. The electrical signal with frequency $\Omega$ is used as the input to an electrical modulator, generating the subcarrier signal. The modulation frequency is chosen to be maximum possible in order to comply with the optical filter rejection bandwidth of 7.5 GHz. The modulation index, controlled by FPGA, is adjusted so that the amplitude ratio between the carrier and all the subcarrier waves is unity. The modulated signal components are then transmitted through the circulator with 0.5 dB losses in each arm to a fiber Bragg grating spectral filter (reflection coefficient of 99.99\% in a 7.5 GHz band, insertion loss 1 dB) with passive thermal stabilization. The optical connections in the detector module introduce additional 1.3 dB loss, so total losses in detection module are 6 dB. Since the sidebands and the carrier follow different optical paths, their losses vary with the transmittance coefficients $\eta_{SB} = 10^{-0.28}$ (2.8 dB loss) and $\eta_{C} = 10^{-0.33}$ (3.3 dB), respectively. The two output ports of the circulator are coupled to the input ports of a self-developed balanced detector (measurement bandwidth 100 MHz, sensitivities are $0.87\pm0.02$ for PD$_1$ and $0.88\pm0.02$ for PD$_2$). Figure \ref{Fig:VarLO} illustrates the variance of the output voltage as a function of the average LO power. The variance increases linearly with increase of the average power, indicating that the quantum noise dominates total variance in the given range power. Experimental measurement outcomes statistics is shown in Fig. \ref{Fig:Noise}. 

\begin{figure}[htbp]
\centering
\includegraphics[width=0.5\linewidth]{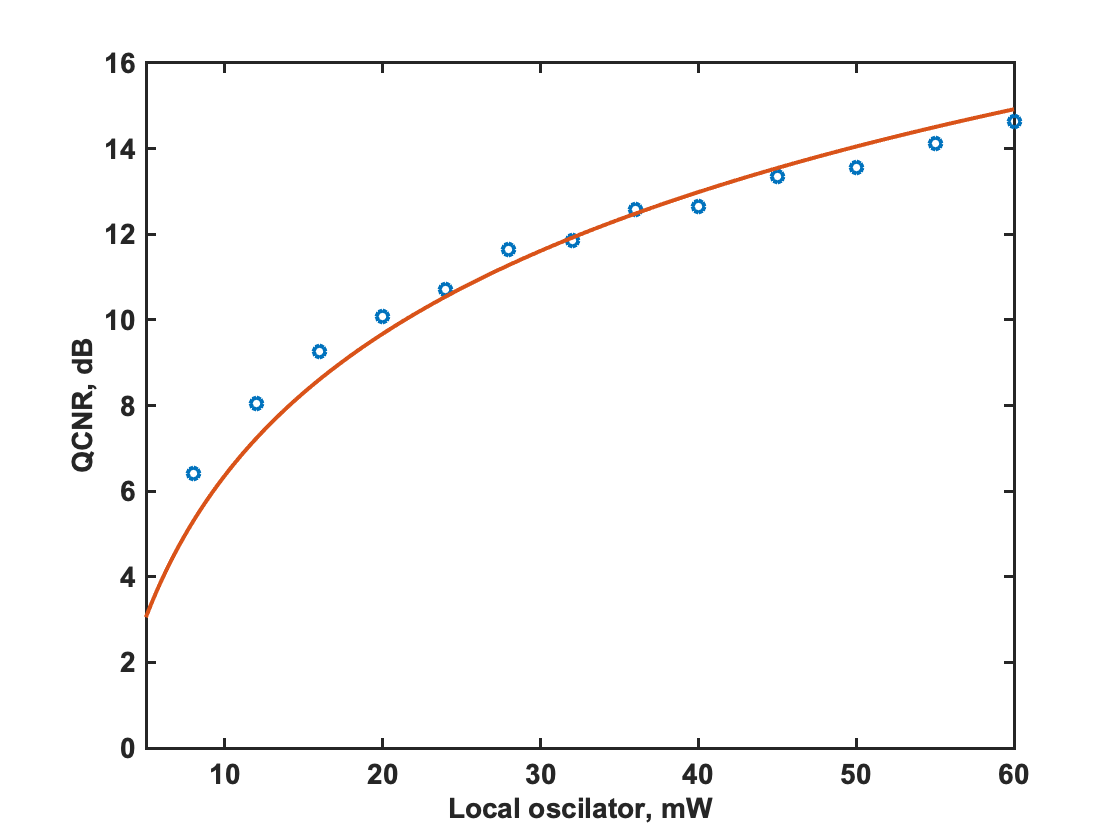}
\caption{Experimental results of QCNR (blue) and its approximation (red). For LO power of 40 mW QCNR is 12.9 dB}
\label{QCNR}
\end{figure}

Since the classical noise, which in theory the eavesdropper can tamper with, always exists in practical QRNGs, it is important to control and eliminate its influence. The measured noise variance can be expressed as:
\begin{equation}
\sigma_m^2 = \sigma_e^2 + \sigma_q^2,
\label{eq:sigma}
\end{equation}
where $\sigma_e^2$ and $\sigma_q^2$ are classical and quantum noise variances, respectively. When LO power is set to 0 mW, the measured voltage variance corresponds to $\sigma_e^2$. In our experiment, it has average measured value of $5.49\times10^{-5} V^2$. Under the same conditions, the average $\sigma_m^2$ value is $1.06\times10^{-3} V^2$ when LO power is set to 40 mW. Thus, the quantum to classical noise ratio (QCNR) is:
\begin{equation}
QCRN = 10 \cdot \log_{10}(\sigma_{q}^{2}/\sigma_{e}^{2}).
\label{eq:QCNR}
\end{equation}
Experimental values of QCNR as a function of LO power are given in Fig.~\ref{QCNR}. 

In the worst case scenario, when the eavesdropped can listen to and control over entire classical noise,   conditional min-entropy parameter can be used to evaluate extractable randomness with a guarantee of security of the random bits \cite{haw2015}. The min-entropy of the discretized measurement signal $M_{dis}$ conditioned on classical noise E is be given by:

\begin{equation}
H_{min} (M_{dis}|E) =\left.-\log_2 \left(\max \left[A,
B)\right]\right) \right.
\label{eq:cond_ent_haw}
\end{equation}

\begin{equation}
A =  \frac{1}{2} \left(erf\left[\frac{e_{max}-R+3\delta/2}{\sqrt{2}\sigma_{q}} \right]+1 \right);  
B = erf\left(\frac{\delta}{2\sqrt{2}\sigma_{q}} \right),
\end{equation}

where R is one half of the \textit{n}-bit ADC input voltage range, $\delta=R/2^{n-1}$ and $e_{max}$ is the maximum value of classical noise outcome. We bound the maximum excursion of classical noise value $e$ as $-5\sigma_{e} \leq e \leq 5\sigma_{e}$. The detector measurement outcomes are sampled by an 8 bit ADC at 100 MSample/s with 0.4 peak-to-peak voltage range. In this case the proposed QRNG has conditional min-entropy of at least 5.85 bits per sample. 

It is important to note that in a practical QRNG the expected value of the measured signal probability distribution can be non-zero. Direct current offset affecting it can be caused by imperfections in the balanced detector, or can be induced by an eavesdropper. As described in \cite{haw2015}, equation \ref{eq:cond_ent_haw} can be generalized to take into account direct current offset of the device, $\Delta$. If the mean value of the electrical noise is not equal to zero, its maximal value can be calculated as $e_{max} = 5\times\sigma_e + \left| \Delta\right|$. We calculated that the conditional min-entropy begins to drop if the classical noise value becomes greater than $\left| \Delta\right| = \pm 0.086$ V. In this case, the first value,$A$, in Eq.~\ref{eq:cond_ent_haw} becomes greater than the second one,$B$. Since QRNG proposed in this paper can effectively compensate the deviation between two arms of the balanced detector by controlling the modulation index of electro-optical phase modulator, we can ensure the direct current offset to remain in the interval $[-0.086 V, 0.086 V]$. 

\begin{table}[ht!]
\centering
\begin{tabular}{lll}
\hline
Statistical test                            & p-value & Result  \\ \hline
Birthday Spacings                           & 0.943542 & success \\
The overlapping 5-permutation test          & 0.294847 & success \\
Binary rank test for 31x31 matrices         & 0.766431 & success \\
Binary rank test for 31x32 matrices         & 0.424927 & success \\
Binary rank test for 6x8 matrices.          & 0.151725 & success \\
Bit stream test                             & 0.119900 & success \\
OPSO                                        & 0.228152 & success \\
OQSO                                        & 0.196286 & success \\
DNA                                         & 0.127439 & success \\
The count-the-1's test on a stream of bytes & 0.354376 & success \\
The count-the-1's test for specific bytes   & 0.137865 & success \\
This is a parking lot test                  & 0.227737 & success \\
The minimum distance test                   & 0.138302 & success \\
The 3drspherestest                          & 0.082485 & success \\
The sqeeze test                             & 0.942042 & success \\
The overlapping sums test                   & 0.186080 & success \\
The runs test                               & 0.461516 & success \\
Craps test                                  & 0.843023 & success \\ \hline
\end{tabular}
\caption{Results of DieHard tests applied to the extracted random numbers. The test is successful if all p-values satisfy $0.025 \leqslant$ p $\leqslant 0.975$.}
\label{tab:DieHard}
\end{table}

In our work we implemented randomness extraction by AES-based cryptographic hashing algorithm with an output length of $l$ = 512 bit \cite{aes}. If the hash function converts $k$ input bits to
\begin{equation}
l< k\cdot H_{min}/n - 2\log_2 (1/\varepsilon),
\label{eq:leftoverhashlemma}
\end{equation}
then the output string $l$ will be $\varepsilon$-close to uniformly distributed random string, according to leftover hash lemma \cite{toma}. We set $k$ = 1024 bit so that the security parameter is $\varepsilon<2^{-100}$. 

The real-time unconditional random-number generation rate of our QRNG is 400 Mbit/s. Better performance can be obtained by improving and optimizing the scheme electrical design, as well as a faster randomness extraction algorithm, which is beyond the scope of this paper. We tested the generated random bits using  NIST and DieHard statistical tests. The results are shown in Tables ~\ref{tab:DieHard} and  ~\ref{tab:NIST}. As one can see, our QRNG has passed all the provided tests, which shows that the patterns are not present in the extracted data.

\begin{table}[ht!]
\centering
\begin{tabular}{lll}
\hline
Statistical test        & p-value  & Result  \\ \hline
Frequency               & 0.534146 & success \\
BlockFrequency          & 0.911413 & success \\
CumulativeSums          & 0.260082 & success \\
Runs                    & 0.350485 & success \\
LongestRun              & 0.213309 & success \\
Rank                    & 0.739918 & success \\
FFT                     & 0.122325 & success \\
NonOverlappingTemplate  & 0.132399 & success \\
OverlappingTemplate     & 0.213309 & success \\
Universal               & 0.739918 & success \\
ApproximateEntropy      & 0.534146 & success \\
RandomExcursions        & 0.159146 & success \\
RandomExcursionsVariant & 0.184362 & success \\
Serial                  & 0.260082 & success \\
LinearComplexity        & 0.534146 & success
\end{tabular}
\caption{Results of NIST tests applied to the extracted random numbers. The test is successful if all P-values satisfy $0.025 \leqslant$ p $\leqslant 0.975$, the minimum pass rate for each statistical test is 0.9583.}
\label{tab:NIST}
\end{table}

\section{Conclusion}
\label{conclusion}
We proposed a vacuum-based quantum random number generator using multi-mode coherent states. We built a mathematical model of the proposed detection scheme and created the experimental setup. We performed a proof-of-principle experiment and demonstrated security-proved random number generation with real-time randomness extraction at the rate of 400 Mbit/s. Future research can be focused on improving the QRNG parameters. 

\section*{Acknowledgements}
This work was financially supported by Russian Ministry of Education (Grant No. 2020-0903)

\section*{Disclosures}
The authors declare no conflicts of interest.



\begin{thebibliography}{43}

\bibitem{abellan2014}
C.~Abell{\'a}n, W.~Amaya, M.~Jofre, M.~Curty, A.~Ac{\'\i}n, J.~Capmany,
  V.~Pruneri, and M.~Mitchell.
\newblock Ultra-fast quantum randomness generation by accelerated phase
  diffusion in a pulsed laser diode.
\newblock {\em Optics express}, 22(2):1645--1654, 2014.

\bibitem{avesani2018}
M.~Avesani, D.~G. Marangon, G.~Vallone, and P.~Villoresi.
\newblock Source-device-independent heterodyne-based quantum random number
  generator at 17 gbps.
\newblock {\em Nature communications}, 9(1):1--7, 2018.

\bibitem{brunner}
N.~Brunner, D.~Cavalcanti, S.~Pironio, V.~Scarani, and S.~Wehner.
\newblock Bell nonlocality.
\newblock {\em Rev. Mod. Phys.}, 86:419--478, Apr 2014.

\bibitem{cao2016}
Z.~Cao, H.~Zhou, X.~Yuan, and X.~Ma.
\newblock Source-independent quantum random number generation.
\newblock {\em Physical Review X}, 6(1):011020, 2016.

\bibitem{chistiakov2019}
V.~Chistiakov, A.~Kozubov, A.~Gaidash, A.~Gleim, and G.~Miroshnichenko.
\newblock Feasibility of twin-field quantum key distribution based on
  multi-mode coherent phase-coded states.
\newblock {\em Optics Express}, 27(25):36551--36561, 2019.

\bibitem{Ferrenberg}
A.~M. Ferrenberg, D.~P. Landau, and Y.~J. Wong.
\newblock Monte carlo simulations: Hidden errors from ``good'' random number
  generators.
\newblock {\em Phys. Rev. Lett.}, 69:3382--3384, Dec 1992.

\bibitem{aes}
P.~FIPS.
\newblock 197: Advanced encryption standard (aes).
\newblock {\em National Institute of Standards and Technology}, 26, 2001.

\bibitem{Gabriel2010}
C.~Gabriel, C.~Wittmann, D.~Sych, R.~Dong, W.~Mauerer, U.~L. Andersen,
  C.~Marquardt, and G.~Leuchs.
\newblock {A generator for unique quantum random numbers based on vacuum
  states}.
\newblock {\em Nature Photonics}, 4(10):711--715, 2010.

\bibitem{gennaro2006}
R.~Gennaro.
\newblock Randomness in cryptography.
\newblock {\em IEEE security \& privacy}, 4(2):64--67, 2006.

\bibitem{Tittel2002}
N.~Gisin, G.~Ribordy, W.~Tittel, and H.~Zbinden.
\newblock Quantum cryptography.
\newblock {\em Reviews of modern physics}, 74(1):145, 2002.

\bibitem{gleim2016}
A.~Gleim, V.~Egorov, Y.~V. Nazarov, S.~Smirnov, V.~Chistyakov, O.~Bannik,
  A.~Anisimov, S.~Kynev, A.~Ivanova, R.~Collins, et~al.
\newblock Secure polarization-independent subcarrier quantum key distribution
  in optical fiber channel using bb84 protocol with a strong reference.
\newblock {\em Optics express}, 24(3):2619--2633, 2016.

\bibitem{Guo2010}
H.~Guo, W.~Tang, Y.~Liu, and W.~Wei.
\newblock Truly random number generation based on measurement of phase noise of
  a laser.
\newblock {\em Physical Review E - Statistical, Nonlinear, and Soft Matter
  Physics}, 81(5):1--4, 2010.

\bibitem{Guo2019}
X.~Guo, C.~Cheng, M.~Wu, Q.~Gao, P.~Li, and Y.~Guo.
\newblock {Parallel real-time quantum random number generator}.
\newblock {\em Optics Letters}, 44(22):5566, 2019.

\bibitem{guo2018}
X.~Guo, R.~Liu, P.~Li, C.~Cheng, M.~Wu, and Y.~Guo.
\newblock Enhancing extractable quantum entropy in vacuum-based quantum random
  number generator.
\newblock {\em Entropy}, 20(11):819, 2018.

\bibitem{haw2015}
J.-Y. Haw, S.~Assad, A.~Lance, N.~Ng, V.~Sharma, P.~K. Lam, and T.~Symul.
\newblock Maximization of extractable randomness in a quantum random-number
  generator.
\newblock {\em Physical Review Applied}, 3(5):054004, 2015.

\bibitem{haylock2019}
B.~Haylock, D.~Peace, F.~Lenzini, C.~Weedbrook, and M.~Lobino.
\newblock Multiplexed quantum random number generation.
\newblock {\em Quantum}, 3:141, 2019.

\bibitem{huang2019}
L.~Huang and H.~Zhou.
\newblock Integrated gbps quantum random number generator with real-time
  extraction based on homodyne detection.
\newblock {\em JOSA B}, 36(3):B130--B136, 2019.

\bibitem{jennewein2000}
T.~Jennewein, U.~Achleitner, G.~Weihs, H.~Weinfurter, and A.~Zeilinger.
\newblock A fast and compact quantum random number generator.
\newblock {\em Review of Scientific Instruments}, 71(4):1675--1680, 2000.

\bibitem{johnston2018random}
D.~Johnston.
\newblock {\em Random Number Generators—Principles and Practices: A Guide for
  Engineers and Programmers}.
\newblock De Gruyter, 2018.

\bibitem{kozubov2019finite}
A.~Kozubov, A.~Gaidash, and G.~Miroshnichenko.
\newblock Finite-key security for quantum key distribution systems utilizing
  weak coherent states.
\newblock {\em arXiv preprint arXiv:1903.04371}, 2019.

\bibitem{kynev2017}
S.~Kynev, V.~Chistyakov, S.~Smirnov, K.~Volkova, V.~Egorov, and A.~Gleim.
\newblock Free-space subcarrier wave quantum communication.
\newblock In {\em J. Phys.: Conf. Ser.}, volume 917, page 052003, 2017.

\bibitem{marsaglia}
Marsaglia.
\newblock Diehard test suite.
\newblock {\em Online: http://www. stat. fsu. edu/pub/diehard}, 1998.

\bibitem{martin2015}
A.~Martin, B.~Sanguinetti, C.~C.~W. Lim, R.~Houlmann, and H.~Zbinden.
\newblock Quantum random number generation for 1.25-ghz quantum key
  distribution systems.
\newblock {\em Journal of Lightwave Technology}, 33(13):2855--2859, 2015.

\bibitem{melnik}
K.~Mel’nik, N.~Arslanov, O.~Bannik, L.~Gilyazov, V.~Egorov, A.~Gleim, and
  S.~Moiseev.
\newblock Using a heterodyne detection scheme in a subcarrier wave quantum
  communication system.
\newblock {\em Bulletin of the Russian Academy of Sciences: Physics},
  82(8):1038--1041, 2018.

\bibitem{merolla1999}
J.-M. Merolla, Y.~Mazurenko, J.-P. Goedgebuer, H.~Porte, and W.~T. Rhodes.
\newblock Phase-modulation transmission system for quantum cryptography.
\newblock {\em Optics letters}, 24(2):104--106, 1999.

\bibitem{miroshnichenko2018}
G.~Miroshnichenko, A.~Kozubov, A.~Gaidash, A.~Gleim, and D.~Horoshko.
\newblock Security of subcarrier wave quantum key distribution against the
  collective beam-splitting attack.
\newblock {\em Optics express}, 26(9):11292--11308, 2018.

\bibitem{miroshnichenkoj}
G.~P. Miroshnichenko, A.~D. Kiselev, A.~I. Trifanov, and A.~V. Gleim.
\newblock Algebraic approach to electro-optic modulation of light: exactly
  solvable multimode quantum model.
\newblock {\em JOSA B}, 34(6):1177--1190, 2017.

\bibitem{nisan1994}
N.~Nisan and A.~Wigderson.
\newblock Hardness vs randomness.
\newblock {\em Journal of computer and System Sciences}, 49(2):149--167, 1994.

\bibitem{pironio2010}
S.~Pironio, A.~Ac{\'\i}n, S.~Massar, A.~B. de~La~Giroday, D.~N. Matsukevich,
  P.~Maunz, S.~Olmschenk, D.~Hayes, L.~Luo, T.~A. Manning, et~al.
\newblock Random numbers certified by bell’s theorem.
\newblock {\em Nature}, 464(7291):1021--1024, 2010.

\bibitem{Raffaelli2018}
F.~Raffaelli, G.~Ferranti, D.~H. Mahler, P.~Sibson, J.~E. Kennard,
  A.~Santamato, G.~Sinclair, D.~Bonneau, M.~G. Thompson, and J.~C. Matthews.
\newblock A homodyne detector integrated onto a photonic chip for measuring
  quantum states and generating random numbers.
\newblock {\em Quantum Science and Technology}, 3(2):025003, 2018.

\bibitem{Renner2008}
R.~Renner.
\newblock Security of quantum key distribution.
\newblock {\em International Journal of Quantum Information}, 6(01):1--127,
  2008.

\bibitem{rukhin}
A.~Rukhin, J.~Soto, J.~Nechvatal, M.~Smid, and E.~Barker.
\newblock A statistical test suite for random and pseudorandom number
  generators for cryptographic applications.
\newblock Technical report, Booz-allen and hamilton inc mclean va, 2001.

\bibitem{Samsonov2019}
E.~Samsonov, R.~Goncharov, A.~Gaidash, A.~Kozubov, V.~Egorov, and A.~Gleim.
\newblock Continuous variable subcarrier wave quantum key distribution with
  discrete modulation: mathematical model and finite-key analysis.
\newblock {\em arXiv preprint arXiv:1910.04431}, 2019.

\bibitem{Shi2016}
Y.~Shi, B.~Chng, and C.~Kurtsiefer.
\newblock {Random numbers from vacuum fluctuations}.
\newblock {\em Applied Physics Letters}, 109(4), 2016.

\bibitem{toma}
M.~Tomamichel, C.~Schaffner, A.~Smith, and R.~Renner.
\newblock Leftover hashing against quantum side information.
\newblock {\em IEEE Transactions on Information Theory}, 57(8):5524--5535,
  2011.

\bibitem{Ulam1949}
N.~M.~S. Ulam.
\newblock The monet carlo method.
\newblock {\em Journal of the American Statistical Association},
  44(247):335--341, 1949.

\bibitem{Varshalovich1988}
D.~Varshalovich, A.~Moskalev, and Khersonski.
\newblock {\em Quantum theory of angular momentum}.
\newblock World Scientific, 1988.

\bibitem{xu2019}
B.~Xu, Z.~Chen, Z.~Li, J.~Yang, Q.~Su, W.~Huang, Y.~Zhang, and H.~Guo.
\newblock High speed continuous variable source-independent quantum random
  number generation.
\newblock {\em Quantum Science and Technology}, 4(2):025013, 2019.

\bibitem{xu2012}
F.~Xu, B.~Qi, X.~Ma, H.~Xu, H.~Zheng, and H.-K. Lo.
\newblock Ultrafast quantum random number generation based on quantum phase
  fluctuations.
\newblock {\em Optics express}, 20(11):12366--12377, 2012.

\bibitem{Xu2016}
F.~Xu, J.~H. Shapiro, and F.~N.~C. Wong.
\newblock {Experimental fast quantum random number generation using
  high-dimensional entanglement with entropy monitoring}.
\newblock {\em Optica}, 3(11):1266, 2016.

\bibitem{zheng2019}
Z.~Zheng, Y.~Zhang, W.~Huang, S.~Yu, and H.~Guo.
\newblock 6 gbps real-time optical quantum random number generator based on
  vacuum fluctuation.
\newblock {\em Review of Scientific Instruments}, 90(4):043105, 2019.

\bibitem{zhou2019}
Q.~Zhou, R.~Valivarthi, C.~John, and W.~Tittel.
\newblock Practical quantum random-number generation based on sampling vacuum
  fluctuations.
\newblock {\em Quantum Engineering}, 1(1):e8, 2019.

\bibitem{Zhu2012}
Y.~Zhu, G.~He, and G.~Zeng.
\newblock {Unbiased quantum random number generation based on squeezed vacuum
  state}.
\newblock {\em International Journal of Quantum Information}, 10(1):1--13,
  2012.

\end{thebibliography}

\end{document}